\documentclass[conference,twocolumn, 10pt]{IEEEtran}
\usepackage{hyperref}
\usepackage{float}
\usepackage{stfloats}
\usepackage{xcolor}
\usepackage[english]{babel} 
\usepackage[utf8]{inputenc}
\usepackage[]{graphicx}
\usepackage{caption}
\usepackage{subcaption}
\usepackage{stmaryrd}
\usepackage{mathtools}
\usepackage[T1]{fontenc}
\usepackage{amsmath, amsthm, amssymb}
\usepackage{pgf,tikz}
\usepackage{bm}

\title{A Learning-Based Approach to Address Complexity-Reliability Tradeoff in OS Decoders}
\author{
\IEEEauthorblockN{Baptiste Cavarec, Hasan Basri Celebi, Mats Bengtsson, and Mikael Skoglund}
\IEEEauthorblockA{
School of Electrical Engineering and Computer Science
\\
KTH Royal Institute of Technology, Stockholm, Sweden
}
}

\begin{document}

\maketitle

\begin{abstract}
In this paper, we study the tradeoffs between complexity and reliability for decoding large linear block codes. We show that using artificial neural networks to predict the required order of an ordered statistics based decoder helps in reducing the average complexity and hence the latency of the decoder. We numerically validate the approach through Monte Carlo simulations.
\end{abstract}

 \section{Introduction}
 

With the exponential increase of number of connected devices and hence of traffic in communications systems, there is more and more concern regarding the tradeoff between decoding complexity and reliability.
To this end, Order Statistics (OS) based decoders have been introduced in \cite{412683} as a variable complexity decoder for linear block codes, in the sense that one can decide on the order and hence the associated complexity. 
However, since the complexity of the OS decoder grows exponentially with the desired order, it is then ill suited for complexity constrained scenarios when the order is large~\cite{8877142}. On the other hand, although selecting a higher order yields higher reliability, it may often yield wasting resources when the received noisy codeword can be decoded with lower order~\cite{7510798}.

The idea of using neural networks (NNs) to decode linear codes is not new. Decoding with NNs was popular since the late 80s' and beginning of 90s' \cite{42215,485019,116658,176685}. However, it is long time avoided due to the lack of available off-line training capacities.
In recent days, \cite{DBLP:journals/corr/NachmaniBB16,DBLP:journals/corr/abs-1802-04741,DBLP:journals/corr/GruberCHB17,8617843} addressed the problem of decoding linear codes by the means of a NN. 
There have been mostly two ways presented in the context of NN based decoders. One way is to learn the optimal weights of a Tanner Graph in order to aid the traditional belief propagation algorithm as presented in \cite{DBLP:journals/corr/GruberCHB17}.
The other type of methods use a NN to decode the codeword based on the received signal  \cite{DBLP:journals/corr/abs-1802-04741,8617843}.
The second kind approach, in particular the one presented in \cite{DBLP:journals/corr/abs-1802-04741}, seems structurally inspired by the structure of OS decoders, while struggling to approach the performance of an OS decoder, of order 2, for large codes.

The major challenge with NN based decoders is training the network with very large number of codewords which is exponential in $K$, denoting the number of information bits encoded in a single codeword. 
Hence, in this paper, to improve the mean performance of OS decoders we present a learning-based method, in a grey box approach, to predict the order of the decoder in a signal adaptive manner. 
This method allows us to significantly decrease the mean complexity of the decoder without sacrificing the reliability. 
Such a method additionally allows to detect signals that may need a higher decoding order than the one a complexity constrained decoder allows and therefore prevents wasting resources.

\textit{Notation}: Vectors and matrices are denoted by bold face lower and upper case letters or symbols, respectively. Matrix $ \textbf{I}_K $ stands for the $ K \times K $ identity matrix. Superscript $ ^* $ denotes the conjugate of a complex number. Norm-2 of a vector $ \boldsymbol{x} $ is denoted by $ ||\boldsymbol{x}||_2 $ and $ \mathcal{CN}(0,\sigma^2) $ represents the circularly symmetric complex additive white Gaussian noise (AWGN) with zero mean and $ \sigma^2 $ variance. All logarithms in this paper are to based 2 and $ \odot $, $ \otimes $, and $ \oplus $ represent the element-wise multiplication, binary multiplication and binary addition, respectively.

\section{System Model}

\subsection{Channel model}

We assume the transmission of a length$ -N $ codeword which is an output of a binary linear block encoder over a binary-input AWGN (Bi-AWGN) channel. The linear binary block code is denoted by $\mathcal{C}(N,K,d_{\min})$ where $ d_{\min} $ represents the minimum Hamming distance between any two codewords.

A codeword $\mathbf{c} \in \mathbb{F}_2^N$ is obtained from the $K$ number of information bits $\mathbf{u}\in \mathbb{F}_2^K$ using the generator matrix $\mathbf{G}$ as 
\begin{equation}\label{key}
\mathbf{c}=\mathbf{u} \otimes \mathbf{G} .
\end{equation}
The codeword is mapped to an antipodal vector $\mathbf{x}$ through the following mapping in $\mathbb{R}$, 
\begin{equation}\label{key}
x_i=2 c_i-1,
\end{equation}
where $ x_i \in \{-1,+1\} $ and $ c_i \in \{0,1\} $ are the $ i $th elements of $ \mathbf{x} $ and $ \mathbf{c} $ sequences, respectively. The signal at the output of the channel is obtained as 
\begin{equation}
\mathbf{y}=\mathbf{x} +\mathbf{w} \label{eq:sysModel}
\end{equation}
where $\mathbf{w} $ is the AWGN noise at the receiver where $ w_i \sim \mathcal{N}(0,\sigma^2) $. The average transmitted signal power is $ \mathcal{E} = 1 $ and the signal-to-noise ratio (SNR) at the receiver is equal to $ 1/\sigma^2 $.

The received noisy sequence $ \mathbf{y} $ is then decoded to find the original information bits. The optimum Maximum Likelihood decoder in such a scenario is given by
\begin{align}
c_\text{ML} &= \hspace{-6pt}\underset{ c\in \mathcal{C}, \: \mathbf{x}=2\mathbf{c}-\boldsymbol{1}}{\text{argmin}} \|\mathbf{y}-\mathbf{x}\|_2
\\
&= \hspace{-6pt}\underset{ c\in \mathcal{C}, \: \mathbf{x}=2\mathbf{c}-\boldsymbol{1}}{\text{argmax}}\text{Re} \left\{ \left\langle \mathbf{y}, \mathbf{x} \right\rangle \right\}.\label{eq:MLDecoder}
\end{align}
However, the complexity of this decoder is of size $2^K$ which becomes computationally impractical even for low values of $K$. 

As introduced in \cite{412683}, OS decoders help in reducing the complexity by making soft decisions on a restrained set of codewords. OS decoder provides a universal soft-decision decoder for linear block codes and it can provide near optimal performance. OS decoders can be implemented as follows:
\begin{itemize}
	\item Define the reliability of each received symbol by $|r_i|=|y_i|$, and re-arrange $ \textbf{r} $ into $ \textbf{r}' $ such that $|r'_1|>|r'_2|>\dots>|r'_N|$.
	
	\item Such arrangement induces a permutation function $\lambda_1[\cdot]$, that also permutes the columns of the code generator matrix by $\mathbf{G}_1=\lambda_1\left[\mathbf{G}\right]$.
	
	\item From $\mathbf{G}_1$, we can create an equivalent matrix $\mathbf{G}_2$ for which the first $K$ columns are independent. This creates a second permutation $\lambda_2$ and an equivalent sequence $\mathbf{z}=\lambda_2[\lambda_1[\mathbf{y}]]$. This construction can be made while ensuring that $|z_1|>|z_2|>\dots>|z_K|$ and $|z_{K+1}|>\dots>|z_N|$. 
	
	\item Moreover, with Gauss-Jordan elimination, $\mathbf{G}_2$ can be converted in systematic form leading to the equivalent generator matrix  
	\begin{equation}\label{key}
	\mathbf{G}_\text{sys}=\left[\mathbf{I}_K~\mathbf{P}\right]
	\end{equation}
	and the associated code $\mathcal{C}_{\text{sys}}$ for which codewords $\mathbf{c}_\text{sys}$, $\exists\mathbf{c} \in \mathcal{C}$ such that $\mathbf{c}_\text{sys}=\lambda_2[\lambda_1[\mathbf{c}]]$.
\end{itemize}

\subsection{Hard Decoding ($0^{th}$ Order Statistics based decoder)}

Given $\mathbf{z}$, defined by the above construction, one can obtain the $0^{th}$ order decoded codeword $ \mathbf{c}^0 $ as the following. Hard decode the most reliable $ K $ bits of $ \textbf{z} $ 
\begin{equation}
z_i^b=\frac{\text{sign}(z_i)+1}{2}, ~~ \text{for} ~0<i<K ,
\end{equation} 
and create the new systematic codeword $\mathbf{a}^0$ as follows
\begin{equation}\label{key}
\mathbf{a}^0=\mathbf{z}^b \otimes \mathbf{G}_\text{sys} .
\end{equation}
Then the output of the $ 0 $th order OSD is
\begin{equation}\label{key}
\mathbf{c}^0=\lambda_1^{-1}\left[ \lambda_2^{-1}[\mathbf{a}^0] \right] .
\end{equation}


\subsection{ Order-$l$ reprocessing}

Based on the decreasing error probability with decreasing indices,  \cite{412683} introduces the concept of order$-l$ reprocessing. Set the list of test error patterns (TEPs), denoted by $ V $, such that it includes all possible length$ -k $ binary sequences with Hamming distance less than or equal to $ l $ and search over the list to find the error sequence that maximizes the likelihood of the systematic codeword to $ \textbf{z} $. This can be formulated as
\begin{align}
\mathbf{a}^l &= \underset{ \left\lbrace \textbf{a}: \: \textbf{a} = (\textbf{z}^b \oplus \textbf{v}) \otimes \textbf{G}_\text{sys}, \: \textbf{v}\in V \right\rbrace }{\text{argmax}}  \mathbb{P}(\mathbf{a}|\textbf{z})
\\
&= \underset{ \left\lbrace \textbf{a}: \: \textbf{a} = (\textbf{z}^b \oplus \textbf{v}) \otimes \textbf{G}_\text{sys}, \: \textbf{v}\in V \right\rbrace }{\text{argmin}}  \| \mathbf{z} - \mathbf{a} \|_2
\label{eq:MLDecoder}
\end{align}
where $\tilde{h}_i=|h_{\lambda_2\circ\lambda_1(i)}|^2$.
The output of the order$ -l $ decoder is 
\begin{equation}\label{key}
\mathbf{c}_\text{OSD}=\lambda_1^{-1}\left[ \lambda_2^{-1}[\mathbf{a}^l] \right]
\end{equation}

Note that \eqref{eq:MLDecoder} in turn yields to comparing all the binary flips of up to $l$ bits of $\mathbf{z}^b$ and obtaining a systematic codeword $\mathbf{a}$ to compute the Euclidean distance to $ \mathbf{z} $ and selecting the one that minimizes the distance.

The cardinality of TEP for an order$ -l $ OS decoder is
\begin{equation}\label{key}
|V| = \sum_{i=0}^{l} {K\choose i}. 
\end{equation}
If $ l = K $, then $ |V| = 2^K $ and all possible codeword comparisons will be taken into account and hence  performance and complexity of the OS decoder will be identical to ML decoder. It is shown in \cite{412683} that if 
\begin{equation}\label{eq_optimal_order}
l \geq \min \left\lbrace \left\lceil \frac{d_{\min}}{4} - 1 \right\rceil, k \right\rbrace ,
\end{equation}
an order$ -l $ OS decoder is asymptotically optimum and near ML performance can be achieved. Thus, $ 2^K - \sum_{i=0}^{l} {K\choose i} $ number of unnecessary codeword comparisons are saved.

\begin{figure*}[t]
	\centering
	\includegraphics[width=0.9\textwidth]{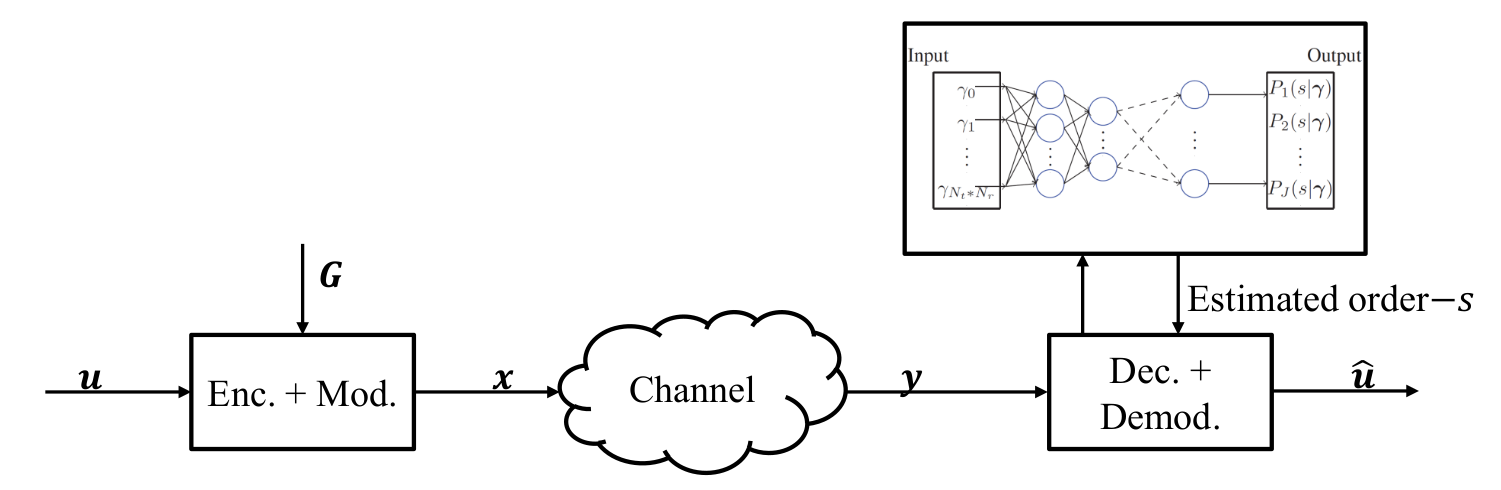}
	\caption{Illustration of the proposed method.} 
	\label{fig:system_model}
\end{figure*}

\section{Complexity-Performance Trade-off}

For a fixed order$ -l $ OS decoder, one needs to compute $\mathbf{G}_\text{sys}$, $\sum_{i=1}^{I} {N \choose i}$ number of new systematic codewords and their subsequent distances. Such procedure increases the complexity exponentially with the desired order while decreasing the error rate of the decoder. The underlying complexity of building $\mathbf{G}_\text{sys}$ and order$ -l $ reprocessing, which are the two main computation intensive operations, are $O \left( N\min\{K,N-K\}^2 \right)$ and $ O \left( NK^{l+1} \right) $, respectively \cite{8877142}. Therefore, for $  l \leq 2 $ the computational complexity is dominated by the formation of $ \mathbf{G}_\text{sys} $, otherwise order$ -l $ reprocessing dominates the complexity. Therefore, order$ -l $ selection allows a complexity-performance trade-off since  a direct relation between complexity and performance is expected for a decoder.

Order$ -l $ selection allows the complexity-reliability tradeoff since a direct relation between complexity and reliability is expected. Given a latency constraint, the corresponding optimal order would be the maximum possible $l$ such that decoding finishes within the time limit, denoted as $l_m$. However, the following two scenarios yield waste of resource/latency on decoding: if (\textit{i}) $l_m < l_r$ (no successful decoding) or (\textit{ii}) $l_m > l_r$ (successful decoding but still wasting resources), where $l_r$ represents the required order$-l$ for decoding $\mathbf{z}$.

Hence, the question that arises in such a scenario is then: \textit{Given a linear code $\mathcal{C}(N,K,d_{\min})$ transmitted over an AWGN channel, is there a way to find the optimal decoding order of the OS decoder for a given received signal $y$?}

\subsection{Baseline: Bound-Based Approach}
\label{sec:Baseline}
When $E_b/N_0$ is known, one may use the tight and relatively simple to compute bounds of \cite{7510798} in order to obtain a guaranteed code error rate for a specific $E_b/N_0$ and order~$l$. Thus by computing such bounds we are able to select the minimum order $l$ guaranteeing the desired performance. This approach has the advantage of providing a fast way of obtaining a \emph{guaranteed} mean performance with a complexity reduction as $E_b/N_0$ increases. However, this method does not apply when the SNR is unknown, one has to estimate the SNR at the receiver in order to be able to use it.

It is to be noted that recently, \cite{yue2020revisit} proposed an approximation for the probability of error of decoding with a given order $l$ based on the knowledge of the received signal and the weight enumerators of the code. They derived decoders according to their approximations, see \cite[Section VII]{yue2020revisit}, however due to the amount of integrals to compute offline, their results could not be reproduced in this paper. The NN based methods presented in the next subsection avoids these computations by directly estimating the probability of success of a given order.
 
\subsection{Learning-Based Approach}

We aim at estimating $l_r$ to successfully decode the codeword. Such task can be seen either as a classification task or a prediction task for a variable representing the success of the order$-l$ OS decoder. The first option is to use a NN as a classifier, that tries to find the class in the sense "what is the minimal order$-l$ that allows to successfully decode the message at the input", by minimizing the cross entropy between the output and the vector representing the class (that has a $1$ at position $l$ and $0$s elsewhere).
The classifier will then tend to output the probability that order $i$ is the exact required order to decode the message at the input, without indicating if the considered order will still yield in a successful decoding, i.e. it outputs $P(\text{"optimal decoding order"}=i)$.

To address the mentioned problem of finding $P(\text{"successfully decoding at order } i\text{"})$, the second option is to have constructions similar to the ones of \cite{NNBayes,8461864}, that may provide guarantees by fixing a threshold on the estimated probability of success such that the predicted order is obtained as in the following.
The idea behind such structures is that by minimizing the cross-entropy between the output of the network and the vector of successes (i.e. the vector that has a $1$ at index $i$ if the order$-l$ decoding of the message is successful and $0$ otherwise) the NN output will approximate the likelihood of successfully decoding with order-$l$ for the given input. 
Further explanations and in depth derivations can be found in \cite{NNBayes}.  
Then, to use this network as an order predictor, set a threshold $\tau\in[0,1]$, given the output of the NN $f_i(\mathbf{z},\mathbf{a}^0)$ for $i \in \{0,1,\dots,l_m\}$ trained to approximate the probability of success of OS decoder of order $i$. The estimated decoding order is obtained by 
\begin{equation}\label{key}
\hat{l}=\arg\hspace{-15pt}\min_{i \in \{0,1,\dots,l_m\}}\hspace{-10pt} \{i|f_i(\mathbf{z},\mathbf{a}^0)\geq \tau\} .
\end{equation}
The appropriate threshold will be determined by the desired codeword error level for the system. 

Beyond the scope of this paper, one can envision using an OS decoder in combination with a NN as presented in here for fading channels other than AWGN, with a slight modification of the OS decoder\footnote{Namely using the reliabilities $r_i=|h_i^*y_i|$ instead of $r_i=|y_i|$} and get similar performance whereas the methods of \cite{7510798,yue2020revisit} would require computing new integrals for each channel realization.
\section{Numerical validation}

In this paper, we built a fully connected feed-forward NN, made of $3$ layers of size $[N,K,l_m+1]$ where $l_m$ denotes the maximum allowed order (due to the constraints), to estimate the order, between $0$ and $l_m,$ necessary to decode the given received signal $\mathbf{z}$ and the hard decoded codeword $\mathbf{a}_0$. A visual illustration of the proposed model is depicted in Fig. \ref{fig:system_model}. 
Note that from \cite{yue2020revisit}, one can infer that inputting the ordered list of reliabilities $|z_i|$ would be sufficient but we obtained worse performance with such structure.

The networks were trained with categorical cross entropy loss for the classification procedure and the mean of each of the output to target cross entropies for the second type of constructions.
Since the scope of this paper is a proof of concept rather than finding the optimal neural network, optimization was conducted using the Adam optimizer \cite{kingma2017adam} with a learning rate of $0.01$.

The training data consisted in $80000$ points, that is the pair (input,output) where the input is the combination $(\mathbf{z},\mathbf{a}_0)$ and the output either the associated optimal decoder order $l^*$ for the classification task or the vector of successes for the success prediction task.
Such a small number of training point is what allows this grey-box approach to work for a relatively large code size when other methods such as \cite{DBLP:journals/corr/abs-1802-04741,DBLP:journals/corr/GruberCHB17} fail because they require $2^K$ or $2^{N-K}$, here $1.844\times 10^{19}$, data points to be trained on.
The training data was generated across 8 different SNRs ranging from $-3$~dB to $4$~dB. 
The impact of the training size and which SNR points to pick were not part of the study.

\begin{figure}
    \centering
        \includegraphics[width=1\columnwidth]{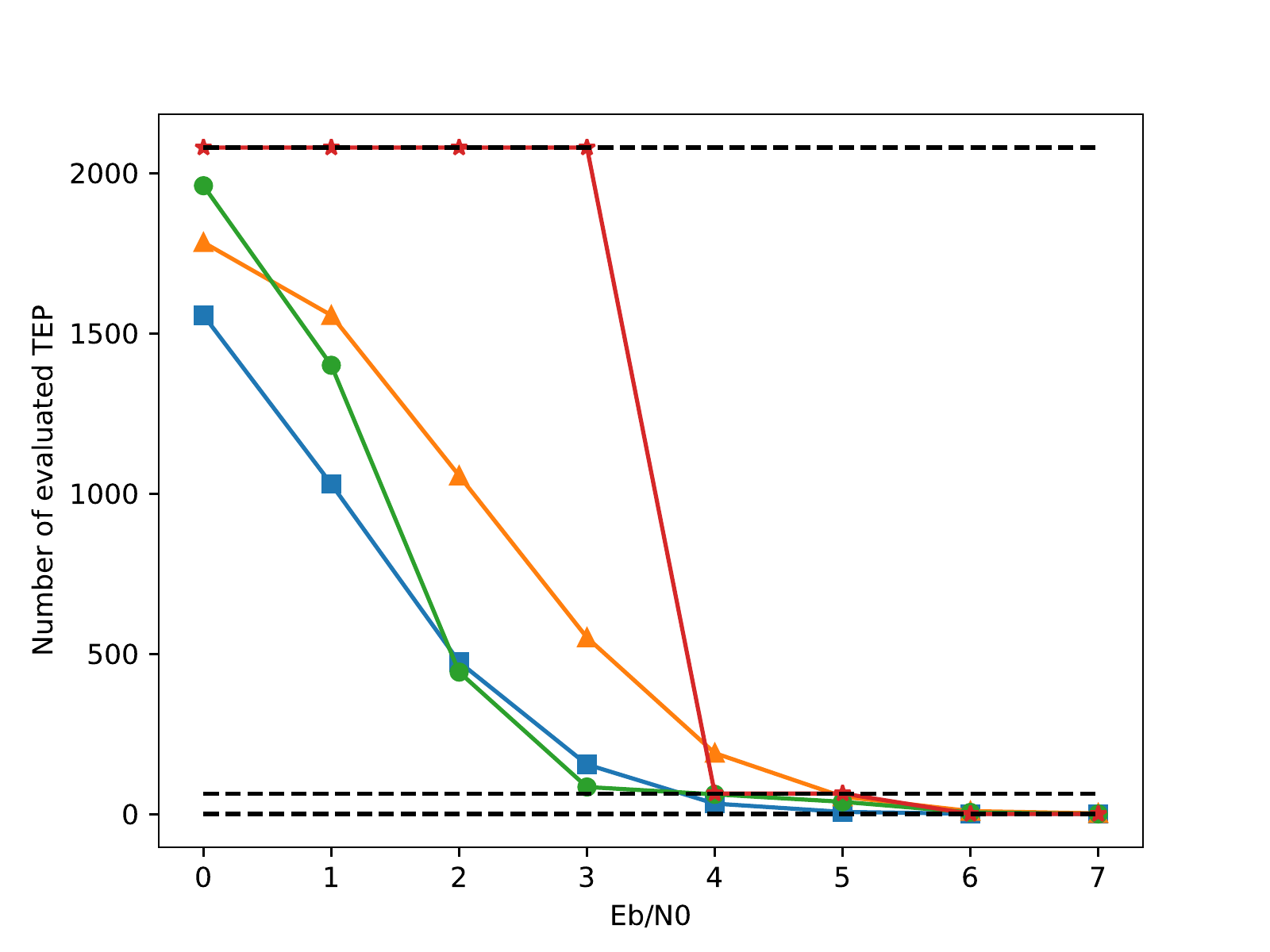} 
        \caption{Comparison of the number of evaluated TEPs (i.e reconfigurations) for $ N=128 $ and $ K = 64 $ (eBCH(128,64,22)), between the bound based baseline when $E_b/N_0$ is known (red), bound based baseline when $E_b/N_0$ is estimated (green), and learning-based classifier NN approach (blue) and  learning-based threshold approach (orange). Reference lines (dashed black) represent the number of evaluated TEPs of the regular OS decoder (orders from 0 to 2).}
        \label{fig:AWGN}
\end{figure}

\begin{figure}
    \centering
           \includegraphics[width=1.0\columnwidth]{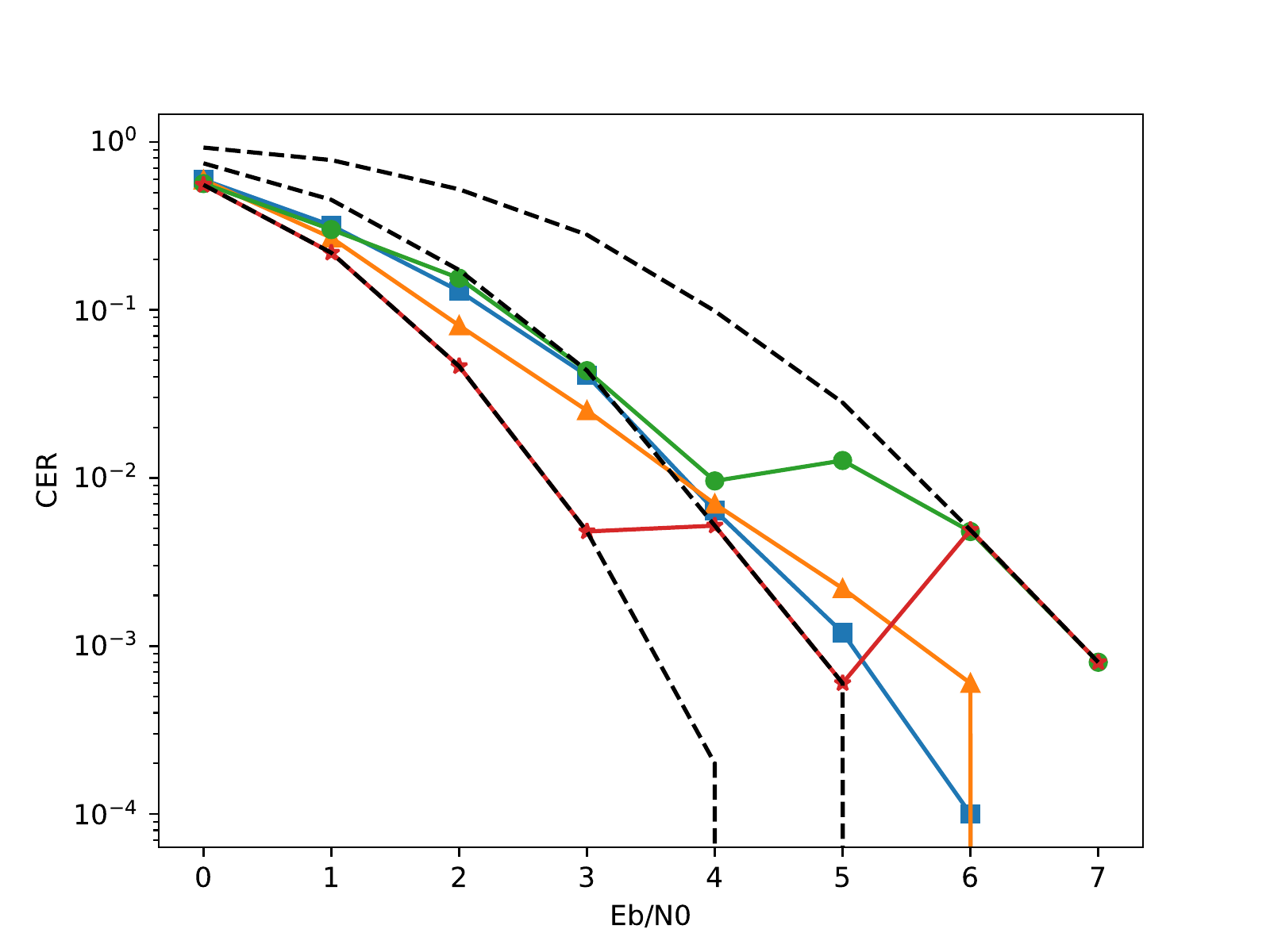} 
        \caption{Comparison of the code error rates for $ N=128 $ and $ K = 64 $ (eBCH(128,64,22)). Please refer to figure~\ref{fig:AWGN} for the information on color codes.}
        \label{fig:AWGNCER}
\end{figure}

Figures~\ref{fig:AWGN} and \ref{fig:AWGNCER} illustrate that for a given latency budget, when the SNR is known the baseline based on the bounds of \cite{7510798} (set with a CER target of $10^{-2}$) provides an optimal complexity performance tradeoff. However, when the SNR is unknown at the receiver (or rather estimated at the receiver), the classifier based NN based approach outperforms the bound based approach, providing a better code error rate with similar latency performance, while the threshold based approach mimics the bound based approach with known SNR while performing better than mere SNR estimation.

\section{Conclusions}

We have shown that neural networks can be used as an order predictor for OS decoders when properly trained. 
The impact on the neural network size, the choice of the training points and the size of the datasets to train on were not studied in this paper but represent a potential area of refinment of this work,
It is left to study whether these techniques can extend to channels beyond AWGN.

\bibliographystyle{IEEEtran}
\bibliography{main}

\end{document}